\documentclass[twocolumn,english,prl]{revtex4}
\usepackage[T1]{fontenc}
\usepackage[latin9]{inputenc}
\usepackage{color}
\usepackage{graphicx}

\makeatletter

\newcommand{\lyxdot}{.}

\@ifundefined{definecolor}
 {\usepackage{color}}{}
\@ifundefined{definecolor}
 {\@ifundefined{definecolor}
 {\usepackage{color}}{}
}{}

\@ifundefined{definecolor}
 {\@ifundefined{definecolor}
 {\usepackage{color}}{}
}{}

\usepackage{babel}

\makeatother

\usepackage{babel}

\makeatother

\usepackage{babel}

\begin{document}

\title{Synchronous Quantum Memories with Time-symmetric Pulses}

\author{Q. Y. He}

\author{M. D. Reid}

\author{P. D. Drummond$^{*}$}

\affiliation{Centre for Atom Optics and Ultrafast Spectroscopy, Swinburne University,
Melbourne VIC 3122, Australia }

\affiliation{$^{*}$Email: pdrummond@swin.edu.au}
\begin{abstract}
We propose a dynamical approach to quantum memories using a synchronous
oscillator-cavity model, in which the coupling is shaped in time to
provide the optimum interface to a symmetric input pulse. This overcomes
the known difficulties of achieving high quantum input-output fidelity
with storage times long compared to the input signal duration. Our
generic model is applicable to any linear storage medium ranging from
a superconducting device to an atomic medium. We show that with temporal
modulation of coupling and/or detuning, it is possible to mode-match
to time-symmetric pulses that have identical pulse shapes on input
and output. 
\end{abstract}
\maketitle
Quantum memories (QM) are key devices both for quantum information
and fundamental tests of quantum mechanics. A QM will write, store
then retrieve a quantum state after an arbitrary length of time. QM
devices are considered vital for the implementation of quantum networks,
quantum cryptography and quantum computing. At a more fundamental
level, they could enable one to generate an entangled quantum state
in one device, then test its decoherence properties in a different
location. This would allow one to test the equivalence of the quantum
state description for more than one physical environment. For example,
there are proposals that gravitational decoherence may occur beyond
the standard model of quantum measurement theory \cite{Marshall2003}.
This would be testable with controlled ways to input, store, then
readout a quantum state in differing environments.

The benchmarks for a QM are storage time and input-output fidelity.
The memory time $T$ must be longer than the duration $T_{I}$ of
the input signal: $T>T_{I}$. Otherwise, the memory is more like a
phase-shifter than a memory. The final quantum state must also be
a close replica of the original. In quantitative terms, the mean state
overlap \cite{Nielsen2000} between the intended and achieved quantum
states (the mean fidelity $\bar{F}$) must satisfy $\bar{F}>\bar{F}_{C}$.
Here $\bar{F}_{C}$ is the best mean fidelity obtainable with a classical
measure and regenerate strategy \cite{Duan2001}. Further to this,
an ideal QM protocol must enable numerous sequential quantum logic
operations to be performed, meaning many input-output {}``quantum
states'', carried on ingoing and outgoing pulse waveforms. This means
that the output pulse envelope should be identical to that of the
input.

\begin{figure}
\includegraphics[height=3cm]{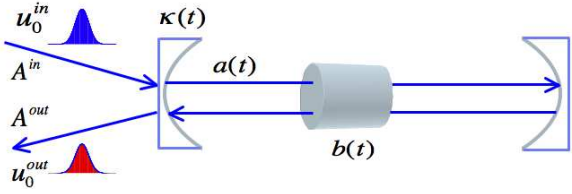}

\caption{(Color online) Proposed dynamical atom-cavity QM. The cavity couples
to only one ingoing and outgoing mode, $u_{0}^{in}$ and $u_{0}^{out}$,
and it is the quantum state of this mode that is stored. The pulse
shape is optimized for efficient writing and reading of the state
onto and from the oscillator medium inside the cavity. A symmetric
pulse shape is used, so the time-reversed output is identical to the
input.\textcolor{blue}{\label{fig:model}}}

\end{figure}

In this paper we propose a new QM protocol (Fig. \ref{fig:model}),
satisfying all of these constraints, in which the {}``state'' is
stored in a dynamically switched cavity-oscillator system. The cavity
acts as an input-output buffer which synchronously mode-matches the
external input pulse to a long-lived internal quantum linear oscillator.
We derive a condition on the time-dependence of the oscillator-cavity
coupling required to match to any external pulse-shape, including
time-symmetric pulses. This contrasts with previous work, in which
the coupling was a step function\cite{He}, resulting in non-symmetric
pulses having different shapes on input and output 

An essential feature of our treatment is that we show how a smooth,
time-symmetric sech-pulse can be stored for times longer than $T_{I}$,
and recalled with high quantum fidelity. This means that the output
pulse-shape replicates the input pulse. Hence, this type of quantum
logic can be cascaded, with interchangeable inputs and outputs. This
is a vital feature of all logic devices. Importantly, we do not use
a slowly-varying pulse approximation \cite{scottkimble2000,lulinprl2000,cavity1Dantan2005,cavity2Dantan2006,cavity8Gorshkov2007,Kalachev2008},
as was required in earlier proposals. This is essential, to allow
the use of fast pulses which can be stored for times much longer than
the pulse-duration.

Our theoretical calculations are carried out with simple non-saturating
linear oscillator models that are analytically soluble. Crucially,
this allows us to calculate pulse-shapes that are dynamically matched
in time to the cavity-oscillator system. This strategy can be combined
with a variety of other technologies. These include quantum nondemolition
(QND) \cite{QMexQNDJulsgaard2004,QMexpQNDMuschik2006,QMexQNDsherson2006,QMexpQNDFiurasek2006}
interactions, Raman and electromagnetically induced transparency (EIT)
\cite{QMexpEITChaneliere2005,QMexpEITChoi2008,QMexpEITEisaman2005,QMexpEITFleischhauer2005,QMexpEITKuzmich2003},
inhomogeneous broadening (CRIB) \cite{CRIBMoiseev2001,CRIBKraus2006,CRIBexpStaudt2007,CRIBexpSangouard2007,CRIBexpHetet2008a,CRIBexpAlexander2006},
superconducting transmission lines and squids \cite{solidstcircuits2008,nori2005,Castellanos-Beltran2008},
magnetic control with a two-level atom \cite{Hollberg2006-1,Hollberg2006-2},
nanomechanical oscillator storage \cite{cavityharris2008,BoseJacobsKnight1999},
and even intra-cavity BEC devices \cite{Brennecke2008}. This opens
some exciting experimental possibilities, including comparisons of
fidelity in QM devices with different effective masses, as a fundamental
test of decoherence in quantum mechanics.

Previous QM experiments were frequently limited by relatively short
storage times \cite{Appel2008}. Other demonstrations focus on retrieval
efficiency at very high photon number \cite{Phillips2008}. However,
these usually have a very low fidelity, since the fidelity at a fixed
efficiency decreases exponentially with photon number. As a rule,
previous proposals either ignore fidelity, or use criteria only applicable
to special known states, like coherent or squeezed states \cite{Appel2008,Adesso2008,Honda2008}.
It is more useful in both applications and fundamental tests to allow
for arbitrary input states. Our analysis is not restricted to any
class of states, except for an upper bound on the input photon number.

\textbf{\emph{Model}}. The quantum information in a temporal mode
of the propagating single transverse-mode operator field $\hat{A}^{in}(t)$
is first transferred to an internal cavity mode with operator $\hat{a}(t)$,
then written into the oscillator or memory with mode operator $\hat{b}(t)$
up to time $t=0$. Subsequently, the interaction is turned off or
detuned for a controllable storage time $T$. The interaction is switched
on again after time $T$, allowing readout into an outgoing quantum
field $\hat{A}^{out}(t)$ at $t>T$ (Fig. \ref{fig:model}). We treat
quantum information encoded into single propagating modes that are
temporally and spatially mode-matched to the memory device \cite{Drummond1981,Drummond1991}.
Here the relevant input and output mode operators are $\hat{a}{}_{0}^{out(in)}=\int u_{0}^{out(in)\star}(t)\hat{A}^{out(in)}(t)dt$,
where $u_{0}^{out(in)}(t)$ is understood to represent the output
(input) temporal mode shape. 

We use the positive P-representation \cite{DrummondGardiner1980},
in which all operators $\hat{A}^{out(in)},\hat{a}_{0}^{out(in)}\,\hat{a},\,\hat{b}$
are formally replaced by c-numbers $A^{out(in)},a_{0}^{out(in)}\, a,\, b$.
Using input-output theory\cite{CollettGardiner1984}, the resulting
dynamical equations are:\begin{eqnarray}
\dot{a}(t) & = & -(\kappa+i\delta(t))a(t)+g(t)b(t)+\sqrt{2\kappa}A^{in}(t)\nonumber \\
\dot{b}(t) & = & -(\gamma+i\Delta(t))b(t)-g(t)a(t)+\sqrt{2\gamma}B^{in}(t)\,.\label{eq:dynamicalequation}\end{eqnarray}

Here $\kappa$ is the cavity damping (assumed fixed), with detuning
$\delta(t).$ The internal cavity-oscillator coupling is $g(t)$ (assumed
variable), while the damping and detuning of the oscillator are $\gamma,\,\Delta(t)$
respectively, with an oscillator reservoir $B^{in}$. These equations
can be applied to a range of experiments ranging from solid-state
crystals or cold atoms to superconducting cavities or nano-oscillators.
The completeness of the representation allows us to treat any quantum
state or memory protocol. Since the equations are linear, the overall
time-delayed input-output relationship must be given by:

\begin{equation}
a_{o}^{out}=\sqrt{\eta_{M}}a_{o}^{in}+\sqrt{1-\eta_{M}}a^{R}.\label{eq:beam-splitter}\end{equation}
Here an amplitude retrieval efficiency $\eta_{M}$ is introduced for
the time-delayed read-out, and $a^{R}$ represents the overall effects
of the loss reservoirs. For simplicity, all reservoirs are assumed
here to be in the vacuum state, without excess phase or thermal noise. 

Hence, we can solve Eq (\ref{eq:dynamicalequation}) to obtain $\sqrt{\eta_{M}}=a_{o}^{out}/a_{o}^{in}$
by integrating over the positive-P output field $A^{out}$. This is
valid since $a^{R}$ corresponds to a bosonic operator which only
acts on a zero-temperature reservoir, and is therefore equal to zero
for the vacuum state in the positive P-representation.

We will analyse the mode-matching conditions for two different dynamical
models with fixed cavity damping $\kappa$. In order to obtain dynamical
mode-matching we require an outgoing vacuum state for $t<T$. In the
positive P-representation this translates to the simple requirement
that $A^{out}=\sqrt{2\kappa}a-A^{in}=0$, so that $A^{in}=\sqrt{2\kappa}a$.
The two models use strategies of either variable coupling or variable
detuning to switch on and off the couping between the oscillator and
the intra-cavity field. For simplicity, we treat the case of zero
internal damping ($\gamma T<<1$) in the equations, while still including
oscillator damping in the graphs to demonstrate that this effect can
be made small if necessary. With no loss of generality, we consider
units for which $\kappa=1$.

\begin{figure}[h]
\includegraphics[width=0.9\columnwidth]{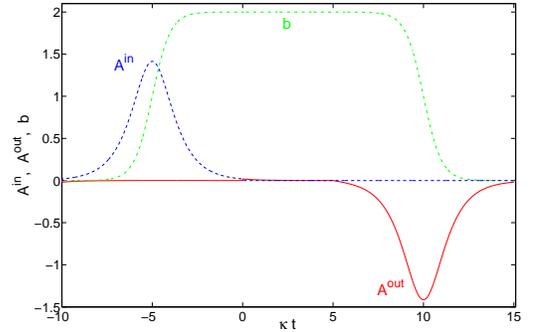}

\caption{(Color online) Case 1: Cavity input (dashed) and output amplitudes
(solid). The dotted line gives the oscillator amplitude. The dashed-dotted
cyan line represents the coupling shape in time $g(t)$. Here $t_{0}=-5$,
$T=5$, $a_{0}=1$. \textcolor{blue}{\label{fig:symmetric coupling shape}}}

\end{figure}

\textbf{\emph{1. Variable coupling}} \textbf{($\Delta,\delta=0$)}.
In this approach, we propose that the cavity decay is fixed, and that
$g(t),$ the interaction of the cavity field with the linear medium,
is switched. During the input stage, the relation $A^{in}=\sqrt{2\kappa}a$
means that $a(t)$ is predetermined for any desired mode-shape $A^{in}(t)$.
This gives an expression for $g(t)$ , since from Eq (\ref{eq:dynamicalequation}),
with $\gamma\rightarrow0$, one has $g(t)=-\dot{b}/a$. Hence:\begin{equation}
[\dot{a}-gb]/a=\dot{a}/a+\dot{(b^{2})}/(2a^{2})=1\,.\label{eq:variablecoupling}\end{equation}

In order to realize a time-symmetric input mode with $a=a_{0}sech(t-t_{0})$,
from Eq (\ref{eq:variablecoupling}) we see that the internal field
amplitude must be $ $$b=a_{0}e^{t-t_{0}}sech(t-t_{0})$.
The optimal shape of the cavity-oscillator coupling in time is therefore
$g(t)=-\dot{b}/a=-sech(t-t_{0})$. This is independent of the amplitude
$a_{0}$ which encodes the quantum information. Here the coupling
is synchronized to $t_{0}$, which is the pulse arrival time. The
quantum memory readout is obtained simply by time-reversal after half
the memory storage time has elapsed, so that $g(t)=g(T-t)$. The resulting
output mode is also time-reversed and is unchanged apart from being
inverted: $A^{out}=-\sqrt{2}a_{0}sech(T+t_{0}-t)$. A typical result
is shown in Fig. \ref{fig:symmetric coupling shape}, from integrating
Eq (\ref{eq:dynamicalequation}).

\begin{figure}[h]
\includegraphics[width=0.9\columnwidth]{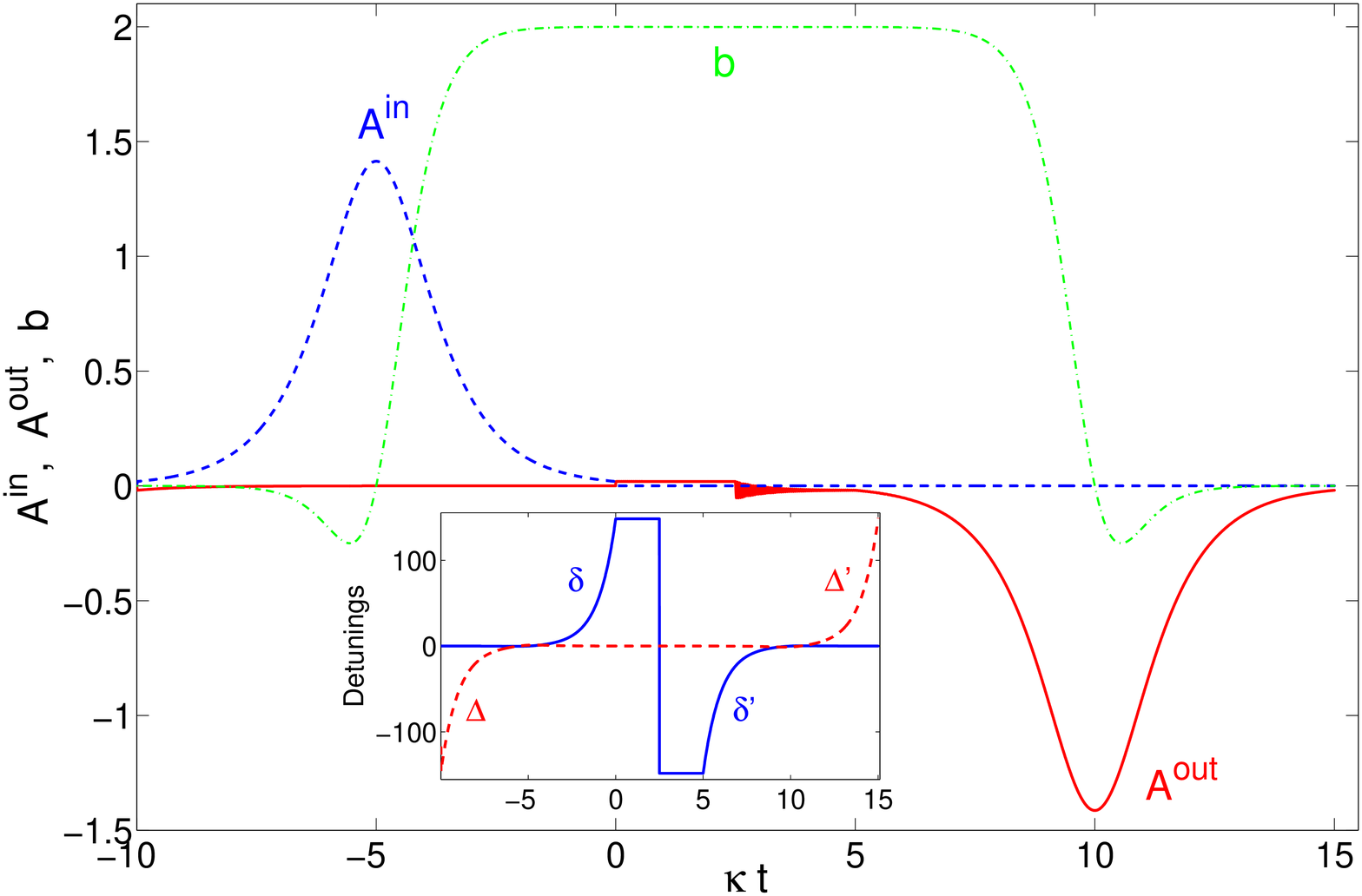}

\caption{(Color online) Case 2: Cavity input (dashed) and output amplitudes
(solid). Other lines and parameters as in Fig (2). The inset gives
the detuning shapes in time: $\Delta(t)$ and $\delta(t)$.\textcolor{blue}{\label{fig:opposite detuning shape}}}

\end{figure}

\textbf{\emph{2. Variable detuning}}\emph{.} In this approach, the
coupling is changed by varying the detunings $\Delta(t)$ and $\delta(t)$.
We consider the simplest case with $g=\kappa=1$ and a symmetric pulse
$a=a_{0}sech(t-t_{0})$. To give a vacuum output during
the writing phase we must have: \begin{eqnarray}
\Delta & = & i(\dot{b}+a)/b\nonumber \\
\delta & = & i(\dot{a}-a-b)/a\label{eq:detunings}\end{eqnarray}
We suppose that $b=b_{1}+ib_{2}$, so that $\delta=i(\dot{a}-a-b_{1})/a+b_{2}/a,$
$\Delta=[(\dot{b}_{1}+a)b_{2}-\dot{b}_{2}b_{1}]/|b|^{2}+i[(\dot{b}_{1}+a)b_{1}+\dot{b}_{2}b_{2}]/|b|^{2}.$
Since $Im(\Delta)=Im(\delta)=0$, we find that $b_{1}=(\dot{a}-a)=-a_{0}e^{t-t_{0}}sech^{2}(t-t_{0})$,
and hence that $b_{2}=a_{0}e^{t-t_{0}}sech(t-t_{0})tanh(t-t_{0}).$
Finally, to realize symmetric input and output pulse shapes, we obtain
from Eq (\ref{eq:detunings}) the required detunings of: $\Delta=e^{-(t-t_{0})}tanh(t-t_{0})+sech(t-t_{0}),$
$\delta=e^{t-t_{0}}tanh(t-t_{0}).$ 

After a controllable storage time, the interaction is switched back
by time reversal of the detunings, so that $\delta'\rightarrow-\delta$
and $\Delta'\rightarrow-\Delta$. The readout is obtained as before,
as shown in Fig. \ref{fig:opposite detuning shape}.

\textbf{\emph{Memory Fidelity}}. Coherent states have proved useful
in quantum applications such as teleportation \cite{Furusawa1998}
and quantum state transfer from light onto atoms \cite{QMexQNDJulsgaard2004}.
It is well known that $\bar{F}_{\bar{n}}^{c}=(1+\overline{n})/(2\overline{n}+1)$
serves as a benchmark for a QM with a gaussian ensemble of coherent
states \cite{Hammerer2005,Braunstein2000} having a mean photon number
$\bar{n}$. For our beam-splitter solution Eq. (\ref{eq:beam-splitter}),
the output for this protocol is $\hat{\rho}_{out}(\alpha)=|\sqrt{\eta_{M}}\alpha\rangle\langle\sqrt{\eta_{M}}\alpha|$,
and the mean fidelity is $\bar{F}_{\bar{n}}=1/[1+\overline{n}(1-\sqrt{\eta_{M}})^{2}]$.
These fidelities may correspond to quite high efficiencies, since
$\eta_{M}>[1-\sqrt{1/(\overline{n}+1)}]^{2}$ is needed for a QM.
With $\overline{n}=20$, QM should be achieved for $\eta_{M}>0.61$,
provided there is no other decoherence.

For many quantum information applications, a larger class of possible
quantum inputs is needed. In the most general case, we can define
the input state as any state with a photon number bounded by $n_{m}$.
This corresponds to an arbitrary state $\left|\vec{\Psi}\right\rangle $
of $1+n_{m}$ levels. $\overline{F}_{n_{m}}$ is then the average
fidelity over all possible coefficients satisfying the constraint
that $\left|\vec{\Psi}\right|=1$. The fidelity limit for (imperfect)
multiple cloning of an arbitrary $1+n_{m}$ level state is $\bar{F}_{n_{m}}^{b}\le2/(n_{m}+2)$
\cite{classicalcloneBuzek1998,classicalcloneHillery1998}. Since a
classical memory can clearly generate any number of copies of a quantum
state, it must be constrained by this fidelity bound also.

We can now calculate the fidelity in the case of $n_{m}=1$ and $n_{m}=2$,
which allows for arbitrary states with up to $1$ and $2$ photons
respectively. After tracing over the reservoir modes, we obtain the
predicted memory fidelities in our beam-splitter model of \cite{He}:
\begin{eqnarray}
\overline{F}_{1} & = & \frac{\eta_{M}+2\sqrt{\eta_{M}}+3}{6}\nonumber \\
\overline{F}_{2} & = & \frac{\eta_{M}^{2}+2\eta_{M}\sqrt{\eta_{M}}+3\eta_{M}+2\sqrt{\eta_{M}}+4}{12}\,.\end{eqnarray}

An arbitrary quantum state fidelity measure gives a better indication
of the power of a QM than a measure constrained to the coherent states.
For example, any storage device with $\eta_{M}>0.23$ can potentially
be a quantum memory for arbitrary states with up to $2$ photons.

\begin{figure}[h]
\includegraphics[width=0.9\columnwidth]{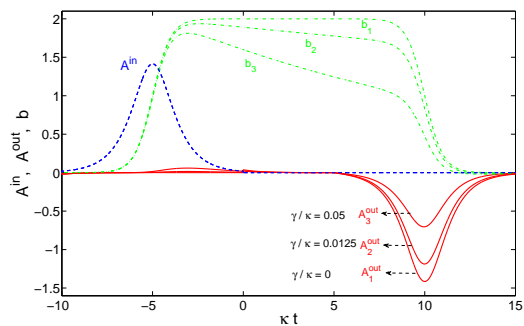}

\caption{(Color online) Case 1 with losses: Cavity input (dashed) and output
amplitudes (solid) with various loss ratios during the storage time{:
$\gamma/\kappa=0,\,0.0125,\,0.05$. Other lines and
parameters as in Fig (2). \label{fig:different oscillator loss}
}}

\end{figure}

\textbf{\emph{Oscillator Loss}}: Figure \ref{fig:different oscillator loss}
shows the typical input-output relation for various loss ratios during
a storage time of duration $T=5$. For $\gamma/\kappa=0,$
$\sqrt{\eta_{M}}=1$, for $\gamma/\kappa=0.0125,$ we
find an efficiency of $\sqrt{\eta_{M}}=0.84$, while for $\gamma/\kappa=0.05,$
we obtain $\sqrt{\eta_{M}}=0.50$. Here we have numerically integrated
Eq (\ref{eq:dynamicalequation}) and used the integral of the mode
overlap with the required sech mode function to obtain the value of
$\sqrt{\eta_{M}}$ from Eq (\ref{eq:beam-splitter}). If $\gamma$
is larger, the oscillator lifetime is shorter, and the information
stored in the medium decays more quickly.

A long storage time $T$ is consistent with high memory fidelity $\bar{F}$,
provided we use dynamical mode matching, and provided $\gamma T\ll1$.
For coherent input states having $\overline{n}=20$, with residual
loss $\gamma/\kappa=0.01$, and storage times $5,$ $10$, $15$,
$20,$ respectively, we find average fidelities $\overline{F}=0.75,$
$0.63$, $0.53$, $0.44$. All except for the last one are larger
than the classical bound $\overline{F}_{20}^{c}=0.51$ required for
a quantum memory. For arbitrary input states of up to two photons,
all these storage times give fidelities larger than the classical
bound $\overline{F}_{2}^{b}=0.5$. Thus, for these parameters, we
are able to predict the existence of a quantum memory with both high
fidelity and relatively long memory lifetime.

In conclusion, we treat a general protocol for a synchronous quantum
memory, using a cavity-oscillator model. We show that with temporal
modulation of coupling and/or detuning, it is possible to mode-match
to identical time-symmetric input and output pulses. Our definition
of an acceptable quantum memory is based on two elementary criteria,
long relative storage times and high quantum fidelity. This type of
quantum memory device promises to satisfy both criteria.

We thank the Australian Research Council for support through ARC Centre
of Excellence and Discovery grants. We grateful to Ping Koy Lam, Elisabeth
Giacobino and others for stimulating discussions.

\bibliographystyle{phaip} \bibliographystyle{phaip} \bibliographystyle{phaip}
\bibliography{QM_database,EPR}

\end{document}